\documentclass[twocolumn, superscriptaddress, preprintnumbers,amsmath,amssymb]{revtex4}
\allowdisplaybreaks
\allowdisplaybreaks[4]

\usepackage{graphicx}% Include figure files
\usepackage{dcolumn}% Align table columns on decimal point
\usepackage{bm}% bold math
\usepackage[dvipdfm,
            pdfstartview=FitH,
            CJKbookmarks=true,
            bookmarksnumbered=true,
            bookmarksopen=true,
            colorlinks=true,
            linkcolor=blue,
            anchorcolor=blue,
            citecolor=blue
            ]{hyperref}

\begin{document}

\title{$g$-factor and static quadrupole moment of $^{135}$Pr,
$^{105}$Pd, and $^{187}$Au in wobbling motion}

\author{C. Broocks} %if needed: \email{catharina.broocks@tum.de}
\affiliation{Physik-Department, Technische Universit\"{a}t
M\"{u}nchen, D-85747 Garching, Germany}

\author{Q. B. Chen}\email{qbchen@pku.edu.cn}
\affiliation{Department of Physics, East China Normal University, Shanghai 200241, China}
\affiliation{Physik-Department, Technische Universit\"{a}t
M\"{u}nchen, D-85747 Garching, Germany}

\author{N. Kaiser}\email{nkaiser@ph.tum.de}
\affiliation{Physik-Department, Technische Universit\"{a}t
M\"{u}nchen, D-85747 Garching, Germany}

\author{Ulf-G. Mei{\ss}ner}\email{meissner@hiskp.uni-bonn.de}
\affiliation{Helmholtz-Institut f\"{u}r Strahlen- und Kernphysik and
Bethe Center for Theoretical Physics, Universit\"{a}t Bonn, D-53115
Bonn, Germany}
\affiliation{Institute for Advanced Simulation, Institut f\"{u}r
Kernphysik, J\"{u}lich Center for Hadron Physics and JARA-HPC,
Forschungszentrum J\"{u}lich, D-52425 J\"{u}lich, Germany}

\date{\today}

\begin{abstract}

The $g$-factor and static quadrupole moment of the nuclides
$^{135}$Pr, $^{105}$Pd, and $^{187}$Au in the wobbling motion are
investigated in the particle-rotor model as functions of the total
spin $I$. The $g$-factor of $^{105}\mathrm{Pd}$ increases with
increasing $I$, due to the negative gyromagnetic ratio of a neutron
valence-neutron. This behavior is in contrast to the decreasing
$g$-factor of the other two nuclides, $^{135}$Pr and $^{187}$Au,
which feature a valence-proton. The static quadrupole moment
$Q$ depends on all three expectation values of the total angular
momentum. It is smaller in the yrast band than in the wobbling band
for the transverse wobblers $^{135}$Pr and $^{105}$Pd, while
larger for the longitudinal wobbler $^{187}$Au.

\end{abstract}

\maketitle

%%%%%%%%%%%%%%%%%%%%%%%%%%%%%%%%%%%%%%%%%%%%%%%%%%%%%%%%%%
%                    begin  introduction
%%%%%%%%%%%%%%%%%%%%%%%%%%%%%%%%%%%%%%%%%%%%%%%%%%%%%%%%%%

The collective wobbling motion of nuclei, as proposed long ago by
Bohr and Mottelson~\cite{Bohr1975}, is a direct evidence for the
existence of nuclides with triaxially deformed shape. Such a
triaxially deformed nucleus rotates about the principal axis
associated to the largest moment of inertia and this axis executes
harmonic oscillations about the space-fixed total angular momentum
vector. The energy spectra related to this collective wobbling
motion consist of a series of rotational bands corresponding to
increasing excitations ($n$) of harmonic oscillation quanta. The
transitions between the rotational bands with $\Delta I=1$ have a
dominant electric quadrupole ($E2$) character.

For the wobbling modes in the presence of a high-$j$ valence-nucleon,
Frauendorf and D\"{o}nau introduced the concepts of \textit{transverse}
wobbling (TW) and \textit{longitudinal} wobbling (LW)~\cite{Frauendorf2014PRC}.
This classification is based on the relative orientation of the angular
momentum of the valence-particle $\bm{j}_p$ and the principal axis
associated to the largest moment of inertia (usually the intermediate
axis). Transverse wobbling means that this relative orientation is
perpendicular, while for longitudinal wobbling it is parallel. The
excitation energy of the wobbling mode with $n=1$ (also called wobbling
energy) in a transverse (longitudinal) wobbler decreases (increases)
with total spin $I$~\cite{Frauendorf2014PRC}.

Experimental data for wobbling bands in odd-mass nuclei are
available for the nuclides
$^{187}\mathrm{Au}$~\cite{Sensharma2020PRL} and
$^{183}\mathrm{Au}$~\cite{Nandi2020PRL} in the $A \approx 190$ mass
region, for $^{161}\mathrm{Lu}$~\cite{Bringel2005EPJA},
$^{163}\mathrm{Lu}$~\cite{Odegaard2001PRL, Jensen2002PRL},
$^{165}\mathrm{Lu}$~\cite{Schonwasser2003PLB},
$^{167}\mathrm{Lu}$~\cite{Amro2003PLB}, and
$^{167}\mathrm{Ta}$~\cite{Hartley2009PRC} in the $A \approx 160$
mass region, for $^{135}\mathrm{Pr}$~\cite{Matta2015PRL,
Sensharma2019PLB}, $^{133}\mathrm{La}$~\cite{Biswas2019EPJA},
$^{130}\mathrm{Ba}$~\cite{Petrache2019PLB, Q.B.Chen2019PRC_v1}, and
$^{127}\mathrm{Xe}$~\cite{Chakraborty2020PLB} with mass number $A
\approx 130$, and moreover for $^{105}\mathrm{Pd}$~\cite{Timar2019PRL}
in the $A \approx 100$ mass region. Among these, $^{133}$La, $^{187}$Au,
and $^{127}$Xe were interpreted as longitudinal wobblers, while
the others are regarded as transverse wobblers.

It should be noted that almost all experiments focus on energy
spectra and electromagnetic transition probabilities, while nuclear
multipole moments are studied very rarely. The $g$-factor and static
quadrupole moment (SQM) have been measured only for the bandhead
state in $^{133}\mathrm{La}$~\cite{Laskar2020PRC} and  their
behavior as a function of spin $I$ were predicted in
Ref.~\cite{Q.B.Chen2020PLB_v1} employing the particle-rotor model
(PRM). At the same time the dependences of the $g$-factor and SQM on
the spin $I$ have been interpreted in detail by analyzing the
angular momentum components of the rotor, proton-particle, and total
nuclear system with the help of various quantum-mechanical
probability distributions. This study has shown that the $g$-factor
and SQM are good indicators of the angular momentum geometry
underlying a triaxially deformed nucleus in  the collective wobbling
motion.

Motivated by these achievements, we will study in the present work
the $g$-factor and SQM for the nuclides $^{135}\mathrm{Pr}$,
$^{105}\mathrm{Pd}$, and $^{187}\mathrm{Au}$ in the collective
wobbling mode. The reason for choosing these three nuclei are that
they lie in different mass regions, and that $^{135}\mathrm{Pr}$ and
$^{187}$Au show the typical TW and LW modes, respectively.
Moreover,  $^{105}$Pd is the first wobbling nucleus with an odd
neutron number.

%%%%%%%%%%%%%%%%%%%%%%%%%%%%%%%%%%%%%%%%%%%%%%%%%%%%%%%%%%
%                    begin  framework
%%%%%%%%%%%%%%%%%%%%%%%%%%%%%%%%%%%%%%%%%%%%%%%%%%%%%%%%%%

Following Ref.~\cite{Q.B.Chen2020PLB_v1}, our present calculations
are carried out within the PRM, which has been used widely and
successfully for describing wobbling bands~\cite{Hamamoto2002PRC,
Hamamoto2003PRC, Frauendorf2014PRC, W.X.Shi2015CPC, Streck2018PRC,
Budaca2018PRC, Q.B.Chen2019PRC_v1, Q.B.Chen2020arXiv, B.Qi2020arXiv,
Q.B.Chen2020arXiv_v1}. The formulas for calculating the $g$-factor and
SQM in PRM can be found in Refs.~\cite{Q.B.Chen2020PLB, Q.B.Chen2020PLB_v1},
but for completeness we give a brief introduction into the formalism.

The $g$-factor is of particular interest, because its value depends
crucially on the alignment of the involved angular momenta, and
moreover it is an experimentally measurable quantity. The $g$-factor
connects the total spin quantum number $I$ with the expectation
value of the magnetic moment $\mu$ and it can be calculated with the
help of the generalized Land\'e formula as:
\begin{align}
 \mu = g (I)I &
     = \frac{\langle II| g_p\bm{j} \cdot \bm{I}
     + g_R \bm{R} \cdot \bm{I} |II\rangle}{I(I+1)}I.
\end{align}
The wave function  $|II\rangle$ refers to $|I, M=I\rangle$ with $M$
the quantum number related to the projection of $\bm{I}$ onto the
$z$-axis in the laboratory frame. The parameters $g_p$ and $g_R$ are
the gyromagnetic ratios of the valence-nucleon (with angular momentum operator
$\bm{j}$) and the collective rotor (with angular momentum operator $\bm{R}$).
Using the relation $ \bm{R}=\bm{I}-\bm{j}$ one can import into the
Land\'e formula some information about the alignment of the involved
angular momenta,
\begin{align}
 g(I) &= \frac{\langle II| g_p\bm{j} \cdot \bm{I}
    + g_R \bm{R} \cdot \bm{I} |II\rangle}{I(I+1)} \label{generalFormulag} \\
   &= g_R + (g_p - g_R) \frac{\langle \bm{j} \cdot \bm{I} \rangle}{I(I+1)} \\
   &= g_R + (g_p - g_R) \frac{j(j+1)}{I(I+1)}
    + (g_p - g_R)\frac{\langle \bm{j} \cdot \bm{R} \rangle}{I(I+1)}.
 \label{gFormulaEndPart1}
\end{align}
where $ \bm{I}=\bm{R}+\bm{j}$ has been applied it the last step. On
the other hand, replacing $\bm{j}$ by $\bm{I}-\bm{R}$ in
Eq.~(\ref{generalFormulag}) gives
\begin{align}
 g(I) = g_p + (g_R - g_p) \frac{\langle \bm{R} \cdot \bm{I} \rangle}{I(I+1)}.
 \label{gFormulaEndPart2}
\end{align}
Combining Eqs.~(\ref{gFormulaEndPart1}) and (\ref{gFormulaEndPart2})
yields an expression for $g(I)$ that depends only on the quantum
numbers $I$, $j$, and the expectation value of the squared rotor
angular momentum $\bm{R}^2$,
\begin{align}
 g(I) = \frac{1}{2} \Big[ (g_p + g_R) + (g_p - g_R) \frac{j(j+1)
      - \langle \bm{R}^2 \rangle}{I(I+1)} \Big].
 \label{gDependingOnR}
\end{align}
If one measures $g(I)$ with its $I$-dependence and knows $j$,
which is the case for a given nucleus,  this formula provides direct
information about the expectation values of $\bm{R}^2$. In the case of
perpendicular or parallel alignment of the angular momenta, simpler
expressions for the $g$-factor are
obtained~\cite{Q.B.Chen2020PLB_v1}:
\begin{align}
 g(I) &= g_p,
   &&   \mathrm{for} \quad \bm{R} \perp \bm{I};
 \label{gforRortI}\\
      &= g_R,
   &&   \mathrm{for} \quad \bm{j} \perp \bm{I};
 \label{gforjortI}\\
      &= g_R + (g_p - g_R) \frac{j(j+1)}{I(I+1)},
   && \mathrm{for} \quad \bm{j} \perp \bm{R};
 \label{gforjortR}\\
      &= g_R + (g_p - g_R) \sqrt{\frac{j(j+1)}{I(I+1)}},
   && \mathrm{for} \quad \bm{j} \parallel \bm{I}.
 \label{gforAllpar}
\end{align}

The other measurable quantity that encodes some information about
the angular momentum geometry is the static quadrupole moment (SQM).
It provides a measure of the non-sphericity of the nuclear charge
distribution and can be calculated as the diagonal matrix-element
\begin{align}
 Q(I)=\langle II|\hat{Q}_{20}|II\rangle,
\end{align}
where the quadrupole moment operator $\hat{Q}_{20}$ in the laboratory frame
is obtained from the intrinsic quadrupole moments $Q_{2\nu}^\prime$ by
multiplication with Wigner $D$-functions:
\begin{align}\label{eq7}
 \hat{Q}_{20}=\sum_{\nu=-2}^2 D_{0,\nu}^{2} Q_{2\nu}^\prime.
\end{align}
The five intrinsic quadrupole moments are $Q_{20}^\prime
=Q_0^\prime\cos\gamma$, $Q_{21}^\prime =Q_{2-1}^\prime=0$, and
$Q_{22}^\prime =Q_{2-2}^\prime=Q_0^\prime\sin\gamma /\sqrt{2}$,
where $Q_0^\prime$ is an empirical quadrupole moment that is related
to the axial deformation parameter $\beta$ by
$Q_0^\prime=3R_0^2Z\beta/\sqrt{5\pi}$, with $Z$ the proton number
and $R_0=1.2\,{\rm fm}\,A^{1/3}$.

One possibility to calculate the SQM is to use the expectation
value of the squared total angular momentum components along the
three principal axes $\langle \hat{I}_k^2 \rangle$~\cite{Q.B.Chen2020PLB,
Q.B.Chen2020PLB_v1},
\begin{align}
 Q(I) &=Q_0(I)+Q_2(I),
 \label{Q0plusQ2Formula}\\
 Q_0(I) &= \frac{3\langle\hat{I}_3^2 \rangle - I(I+1)}{(I+1)(2I+3)} Q_0'\cos{\gamma},
 \label{Q0Formula}\\
 Q_2(I) &= \frac{\sqrt{3} (\langle\hat{I}_1^2 \rangle
 - \langle\hat{I}_2^2 \rangle)}{(I+1)(2I+3)} Q_0'\sin{\gamma}.
 \label{Q2Formula}
\end{align}
This form has the advantage that one can  get some information
about the spin orientation from $Q(I)$.

For the nuclides $^{135}\mathrm{Pr}$~\cite{Matta2015PRL,
Sensharma2019PLB}, $^{105}\mathrm{Pd}$~\cite{Timar2019PRL}, and
$^{187}\mathrm{Au}$~\cite{Sensharma2020PRL}, the PRM has been used
to reproduce the experimental energy spectra and wobbling energies
together with the electromagnetic transition probability ratios,
$B(M1)_{\textrm{out}}/B(E2)_{\textrm{in}}$ and
$B(E2)_{\textrm{out}}/B(E2)_{\textrm{in}}$ for their wobbling bands.
In this work, we focus on the $g$-factor and SQM. The pertinent
numerical details, including the  deformation parameters, moments of
inertia, $g$-factors of valence-nucleon and rotor, and empirical
quadrupole moments, can be found in Refs.~\cite{Matta2015PRL,
Sensharma2019PLB, Timar2019PRL, Sensharma2020PRL}.

%%%%%%%%%%%%%%%%%%%%%%%%%%%%%%%%%%%%%%%%%%%%%%%%%%%%%%%%%%
%                    begin  results and discussion
%%%%%%%%%%%%%%%%%%%%%%%%%%%%%%%%%%%%%%%%%%%%%%%%%%%%%%%%%%

\begin{figure}[ht]
    \centering
    \includegraphics[width=7.0 cm]{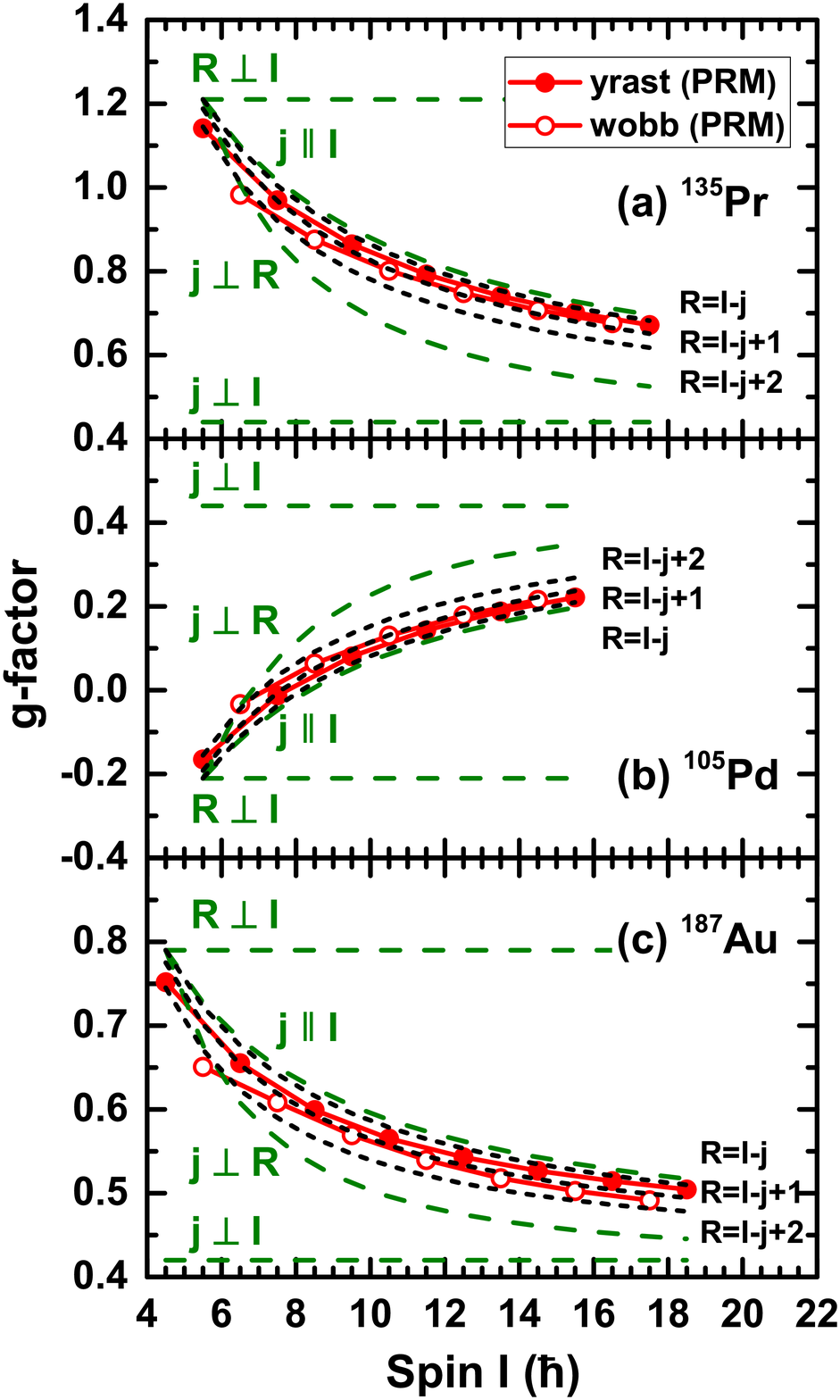}
    \caption{PRM calculation of the $g$-factor in the yrast and wobbling
    band for the nuclei $^{135}\mathrm{Pr}$ (a),
    $^{105}\mathrm{Pd}$ (b), and  $^{187}\mathrm{Au}$ (c).
    The dashed lines for special alignment cases
    $\bm{R} \perp \bm{I}$, $\bm{j} \perp \bm{I}$, $\bm{j} \perp \bm{R}$,
    and $\bm{j} \parallel \bm{I}$ are calculated with
    Eqs.~(\ref{gforRortI})-(\ref{gforAllpar}). The dotted
    curves show the trend of the $g$-factor, approximating the rotor angular
    momentum by $\langle \bm{R}^2 \rangle = R(R+1)$
    with $R=I-j$, $R=I-j+1$ or $R=I-j+2$,  and using Eq.~(\ref{gDependingOnR}).
    \label{gfacGesamt}}
\end{figure}

We will first discuss the results for the $g$-factor of these three wobbling
nuclei. The calculated values of $g(I)$  are shown in Fig.~\ref{gfacGesamt}.
For $^{135}\mathrm{Pr}$ (see Fig.~\ref{gfacGesamt}(a)) the $g$-factor
decreases with increasing $I$ down to a value $g(I=15.5) = 0.7$
and this decrease is faster for small $I$-values. The results for
$^{187}\mathrm{Au}$ displayed in Fig.~\ref{gfacGesamt}(c) show the
same trend with $g(I=4.5)  = 0.75 $ at the bandhead and a
decrease down to $g(I=18.5) = 0.5$. However, the $g$-factor
of $^{105}\mathrm{Pd}$ (see Fig.~\ref{gfacGesamt}(c)) shows an
inverted behavior with an increasing $g$-factor from $g(I=5.5) =
-0.2$ at the bandhead up to $g(I=15.5) = 0.2$ at the highest calculated
$I$-value.

\begin{figure}[ht]
    \centering
    \includegraphics[width=6.0 cm]{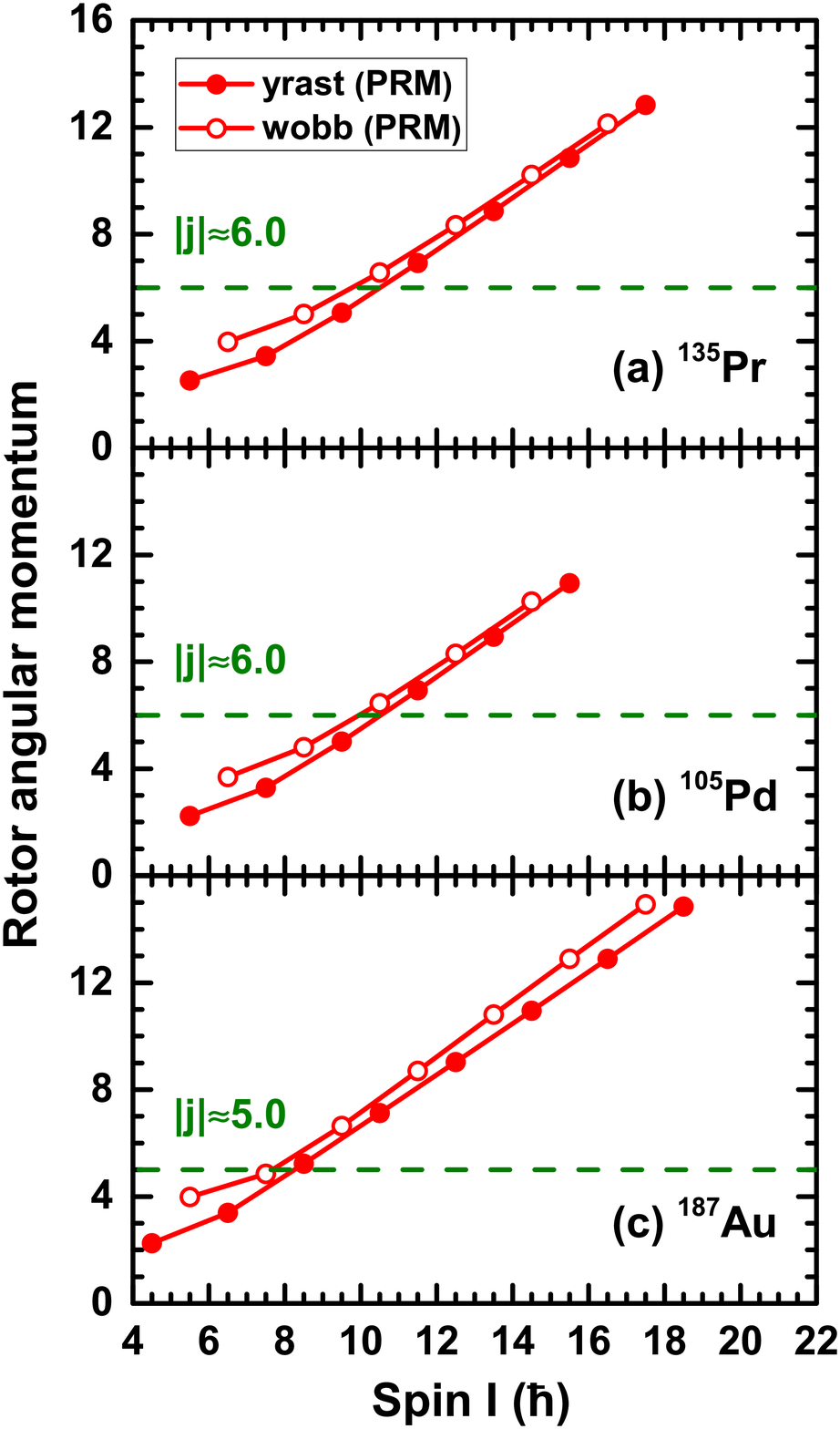}
    \caption{The square root of the expectation value of the squared
    rotor angular momentum $\bar{R}=\langle \bm{R}^2 \rangle^{1/2}$
    for states in the yrast and wobbling bands of $^{135}$Pr, $^{105}$Pd, and
    $^{187}$Au. The horizontal line in each plot corresponds to the
    length of the particle angular momentum $\bar{j} \approx j+0.5$.
    It delineates the region of low spins $I$, where the particle
    angular momentum is larger than the rotor angular momentum.
    \label{fig3}}
\end{figure}

One can explain the behavior of the $g$-factor shown in Fig.~\ref{gfacGesamt}
by considering its dependence on the gyromagnetic ratios $g_p$ and $g_R$,
the quantum numbers $j$ and $I$, and the expectation value of the rotor
angular momentum $\sqrt{\langle \bm{R}^2\rangle}$, as written in
Eq.~(\ref{gDependingOnR}). For this purpose we show in
Fig.~\ref{fig3} the expectation value of the rotor angular momentum
$\bar{R}=\sqrt{\langle \bm{R}^2\rangle}=\sqrt{\langle R_s^2\rangle
+\langle R_m^2\rangle+\langle R_l^2\rangle}$. A comparison of these
plots for the three nuclei reveals that $^{135}\mathrm{Pr}$ and
$^{105}\mathrm{Pd}$ feature very similar $\bar{R}$-values.
For small values of spin $I$, the $\bar{R}$-values
in the wobbling band are larger than in the yrast band, but with
increasing $I$, this difference becomes smaller. One observes that
$\bar{R}$ increases almost linearly from $\bar{R}(I=5.5)=2$ in the
yrast band and $\bar{R}(I=6.5)= 4$ in the wobbling band up to a
similar value of $\bar{R}(I = 18.5) = 15$. The values for
$^{187}\mathrm{Au}$ are slightly higher. While $j$ is constant
for each nucleus, one observes that $\bar{R}$ increases with
growing $I$.

In order to explain the decreasing and increasing $g$-factors, we
take a closer look at both parts in Eq.~(\ref{gDependingOnR}). The
constant offset $(g_p + g_R)/2$ is $ 0.8$ for $^{135}\mathrm{Pr}$
and $ 0.6 $ for $^{187}\mathrm{Au}$, but has a much smaller value of
$ 0.1$ for $^{105}\mathrm{Pd}$. The second part in the $g$-factor formula
is of the form: $(g_p - g_R)[j(j+1)-\langle \bm{R}^2 \rangle]/[2I(I+1)]$.
For the nuclei with $j=5.5$ ($^{135}\mathrm{Pr}$ and
$^{105}\mathrm{Pd}$), $\bar{R}$ is smaller than $\bar{j}=
\sqrt{j(j+1)}\approx 6.0$ up to $I=9.5$. For $^{187}\mathrm{Au}$ with
$j=4.5$, $\bar{R}$ is smaller than $\bar{j}\approx 5.0$ up to
$I=7.5$. Therefore, the difference is positive for small  $I$ and it
becomes smaller or even negative with increasing $I$. Since
the coefficient $g_p-g_R$ is positive for $^{135}\mathrm{Pr}$ and
$^{187}\mathrm{Au}$, the $g$-factors of these two nuclei are bigger
than the constant $(g_p + g_R)/2$ at the bandhead and they decrease
with increasing $I$. Because of the valence-neutron in $^{105}\mathrm{Pd}$,
$g_p-g_R$ is negative and the second piece is negative for
small $I$ and positive for $I>9.5$. This explains the
increasing $g$-factor of this nucleus. In the limit of very high
$I$, the rotor angular momentum tends to $\langle \bm{R}^2 \rangle
\rightarrow I(I+1)$ and $j$ becomes negligible  $j(j+1) \ll
\langle\bm{R}^2 \rangle$. In this case the $g$-factor approaches the
value $g_R$.

Since $g_p$ and $g_R$ are measurable, $\langle \bm{R}^2 \rangle$ is
the only unknown in Eq.~(\ref{gDependingOnR}) and can therefore be
extracted from the experimental $g$-factor for each single
spin-state $I$. Nevertheless, it can be helpful to compare the
measured $g$-factors  with a model calculation when  $\langle
\bm{R}^2\rangle$ is significant in order to see tendencies and to do
a consistency checks. For this reason we insert in Fig.~\ref{gfacGesamt}
three dotted lines that give the $g$-factor obtained with rotor angular
momentum quantum numbers $R=I-j$, $R=I-j+1$, and $R=I-j+2$ using the
relation $\langle \bm{R}^2 \rangle = R(R+1)$. Since all three nuclei have
very similar values of $\langle \bm{R}^2\rangle$, the calculated
$g$-factors tend to the same lines of approximation. At the bandhead
the $g$-factor of the yrast state is close to the $R=I-j+2$ line,
where the rotor angular momentum is approximated by
$\sqrt{R(R+1)}=2.44$. For large values of $I$, the $g$-factors of
yrast and wobbling states are in better agreement with the curves
for $R=I-j$ and $R=I-j+1$, respectively. The slightly larger
$\langle \bm{R}^2 \rangle$-values for $^{187}\mathrm{Au}$ favor the
$R=I-j+2$ curve for the $g$-factors of  wobbling states.

The third set of curves displayed in Fig.~\ref{gfacGesamt} are the
dashed lines corresponding to specific alignment situations
$\bm{R} \perp \bm{I}$, $\bm{j} \perp \bm{I}$, $\bm{j} \perp \bm{R}$,
and $\bm{j} \parallel \bm{I}$. For small $I$ in the yrast band,
the calculated $g$-factor is close to any of the $\bm{R} \perp \bm{I}$,
$\bm{j} \perp \bm{R}$, and  $\bm{j} \parallel \bm{I}$ curves. These curves
intersect at the bandhead since Eqs.~(\ref{gforRortI}),
(\ref{gforjortR}), and (\ref{gforAllpar}) give there the same value $g =
g_p$.

For higher spin $I$, the calculated $g$-factors for all three nuclei
lie between the $\bm{j} \parallel \bm{I}$ and the $\bm{j} \perp
\bm{R}$ curves, tending towards $\bm{j} \parallel \bm{I}$ for
increasing $I$. The $g$-factors in yrast states are always closer to the
parallel-alignment curve than those in wobbling states. This is
plausible, since $\bm{I}$ in a wobbling state is tilted further away
from $\bm{j}$ than in the respective yrast case. For the first
wobbling band this implies that $\bm{R}$ has a better agreement with perpendicular
alignment to $\bm{j}$ than in the case of the yrast band. The increasing
agreement for both bands with the $\bm{j}
\parallel \bm{I}$ curve is a clue for the change from a transverse
to a longitudinal alignment of the particle angular momentum.

\begin{figure}[ht]
    \centering
    \includegraphics[width =7.0 cm]{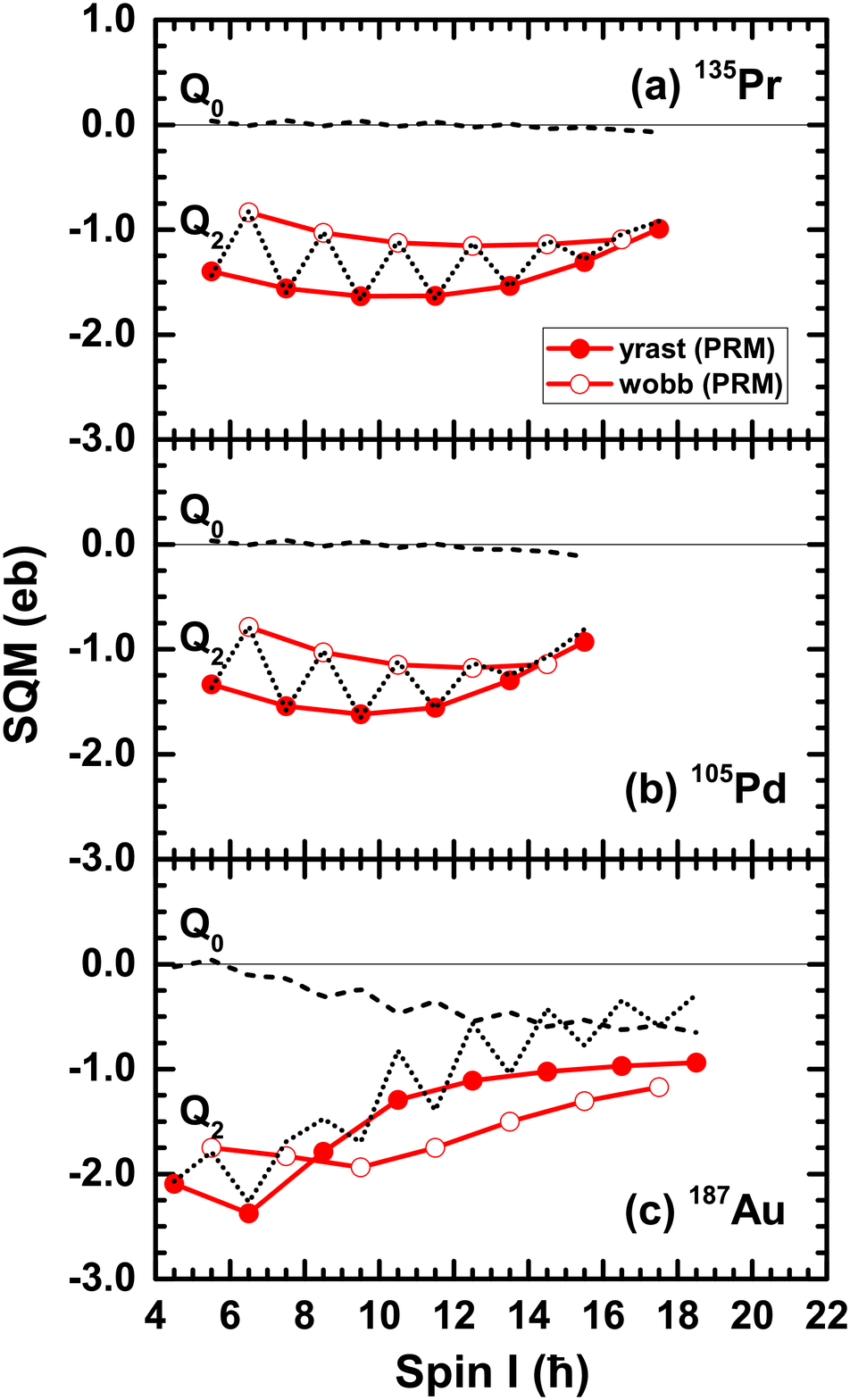}
    \caption{Static quadrupole moments for states in the yrast and wobbling bands of
    the nuclei $^{135}\mathrm{Pr}$(a), $^{105}\mathrm{Pd}$ (b)
    and $^{187}\mathrm{Au}$ (c). The contributions $Q_0$ and $Q_2$ are calculated
    with Eqs.~(\ref{Q0Formula}) and (\ref{Q2Formula}) using the choice $\gamma
    \to \gamma+240^\circ$.}
    \label{SQMGesamt}
\end{figure}

Next, we present in Fig.~\ref{SQMGesamt} the SQM results for the
three nuclei, based on Eq.~(\ref{Q0plusQ2Formula}). In the calculation,
we replace the triaxial deformation parameter $\gamma $ by $ \gamma
+240^\circ$ which corresponds to the same shape, but then $Q_0$ vanishes
almost due the smallness of $\cos(\gamma+240^\circ)=-0.087$ with
$\gamma\sim 25^\circ$ for the three nuclei~\cite{Matta2015PRL,
Sensharma2019PLB, Timar2019PRL, Sensharma2020PRL}. This setup provides a
more promising chance to gain information about the angular momentum
components from the second contribution $Q_2(I)$. Note that according to
this choice the assignment of principal axes is: $1\to s$, $2\to l$,
and $3\to m$. The individual contributions $Q_0(I)$ and $Q_2(I)$ to
$Q(I)$ are displayed in Fig.~\ref{SQMGesamt} by dashed and dotted
lines, respectively.

A first observation is that the SQM is negative for all three
nuclei and every calculated $I$-value, which is attributed to
a much smaller angular momentum component of the $l$-axis
compared to the $s$-axis.

Fig.~\ref{SQMGesamt}(a) shows that the SQM of $^{135}\mathrm{Pr}$ changes
from decreasing to increasing at $I=9.5$ in the yrast band
and a similar trend is indicated at $I=12.5$ in the wobbling
band. However, the change of the SQM in the wobbling band is not
very pronounced and it appears to be almost constant for high spins.
The difference between the SQMs in the yrast and wobbling band decreases
with increasing $I$, and the two curves cross almost at
$Q(I=16.5) =-1.1~e\textrm{b}$. The contribution $Q_0(I)$
has a minor tendency towards negative values, but in
comparison to $Q_2(I)$ it can be approximated by $Q_0(I)=0$.
With respect to the SQMs the nucleus $^{105}\mathrm{Pd}$ shown in
Fig.~\ref{SQMGesamt}(b) exhibits a behavior very similar to
$^{135}\mathrm{Pr}$. At the value $Q(I=14.5) = -1.15~e\textrm{b}$,
the SQM in the yrast band becomes larger than that in the wobbling
band. Note that both nuclides were interpreted as transverse
wobblers at the low spins~\cite{Matta2015PRL, Timar2019PRL,
Sensharma2019PLB}. At higher spins, a transition to longitudinal
wobbling can happen (e.g., in $^{135}$Pr~\cite{Matta2015PRL}
for $I\geq 14.5$). Hence the crossing of SQM curves of yrast
and wobbling bands indicates the transverse wobbling collapses.

In Fig.~\ref{SQMGesamt}(c) a slightly different behavior can be seen
for the nuclide $^{187}\mathrm{Au}$, which according to
Ref.~\cite{Sensharma2020PRL} supports the picture of a longitudinal
wobbler. For states in the yrast band the SQM starts from
$Q(I=4.5)= -2.1~e\textrm{b}$, decreases to $Q(I=6.5) =
-2.4~e\textrm{b}$, and then develops a declining increase up to
$Q(I=18.5) = -1.0~e\textrm{b}$. The SQM of states in the wobbling
band first decreases from $Q(I=5.5) = -1.75~e\textrm{b}$ to
$Q(I=9.5) = -2.0~e\textrm{b}$, and beyond that its value increases
almost linearly up to $Q(I=17.5) = -1.17~e\textrm{b}$. The difference
between SQM-values for states in the yrast and wobbling band decreases
again at high spin $I$. Whereas $Q_0(I)$ vanishes for the low spins,
it becomes even larger in magnitude than $Q_2(I)$ at the higher spins.
This is caused by the longitudinal wobbling motion, in which the
total angular momentum, as shown in the following, aligns along the
$m$-axis.

\begin{figure}[ht]
    \centering
    \includegraphics[width=6.0 cm]{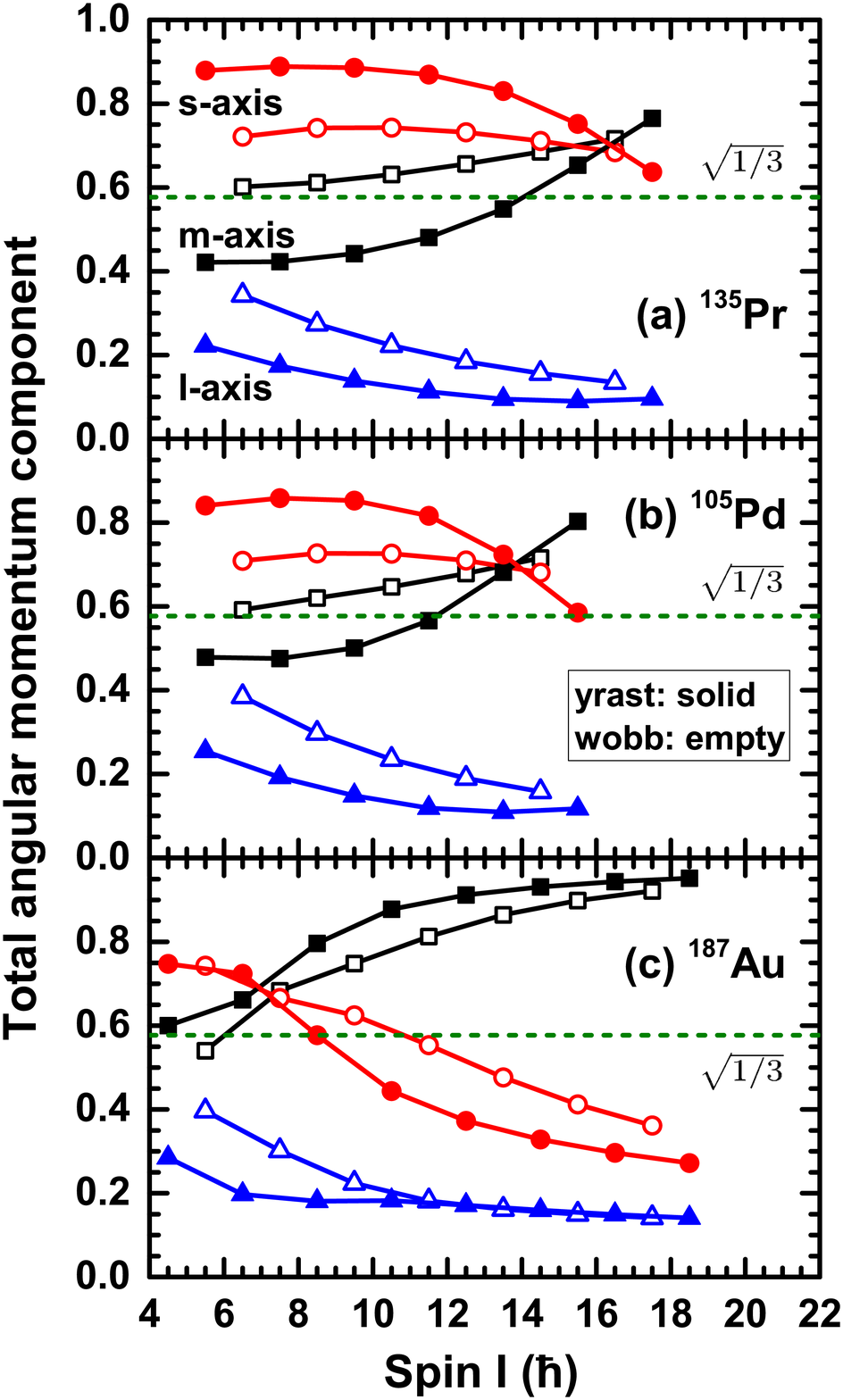}
    \caption{Ratio-plots $r_k=\sqrt{\langle I_k^2\rangle}/\sqrt{I(I+1)}$
    of the total angular momentum components along the short ($s$-),
    intermediate ($m$-), and long ($l$-) axes for states in the
    yrast and wobbling bands of $^{135}$Pr, $^{105}$Pd, and $^{187}$Au.}
    \label{AmAlignPlot}
\end{figure}

For $^{135}\mathrm{Pr}$ and $^{105}\mathrm{Pd}$ the behavior of the
SQM follows from the expression $Q_2(I)$ in Eq.~(\ref{Q0Formula}),
since $\cos\gamma$ is close to zero and/or $3 \langle \hat{I}_m^2
\rangle \approx I(I+1)$. To check the last condition and to show the
behavior of individual total angular momentum components,
ratio-plots
\begin{align}
r_k=\frac{I_k}{\sqrt{I_s^2+I_m^2+I_l^2}}
   =\frac{\sqrt{\langle \hat{I}_k^2\rangle}}{\sqrt{I(I+1)}}
\end{align}
of the total angular momentum components along the short ($s$-),
intermediate ($m$-), and long ($l$-) axes with respect to the length
of the total spin are presented in Fig.~\ref{AmAlignPlot}. The dotted
horizontal line at height $\sqrt{1/3}$ corresponds to the situation
$3 \langle \hat{I}_m^2 \rangle = I(I+1)$.

There is good agreement between $r_m$ and $\sqrt{1/3}$ for states in
the wobbling band of $^{135}\mathrm{Pr}$ and $^{105}\mathrm{Pd}$ at
low spins $I$, as can be seen in Fig.~\ref{AmAlignPlot}. The
decreasing SQM for small $I$ is caused by the denominator
$(I+1)(2I+3)$, whereas for higher $I$ the numerator
$\sqrt{3}(\langle \hat{I}_s^2\rangle -\langle \hat{I}_l^2 \rangle)$
is increasing faster. The $l$-component $I_l$ is almost constant
with $I_l \approx 1.5$ for states in the yrast band and $I_l \approx 2.5$
for states in the wobbling band. The $s$-component $I_s$
is larger for states in the yrast band than
in the wobbling band and it increase with $I$. This is the
reason, why the SQM difference is positive and has larger values in
yrast band than in the wobbling band. The SQM is always negative due
to the negative prefactor $\sin (\gamma+240^\circ)$.
The nuclides $^{135}\mathrm{Pr}$ and $^{105}\mathrm{Pd}$ show the
same tendency.

For $^{187}\mathrm{Au}$, the absolute value of the SQM of the yrast
states becomes smaller than that in the wobbling states for larger $I$,
because $I_s$ of the yrast states gets smaller than $I_s$ of the wobbling
states. This is a feature of a longitudinal wobbler with respect to
the $m$-axis.

To explain the behavior of the SQM for $^{187}\mathrm{Au}$ more
detailed, one needs to consider the $Q_2(I)$-part of $Q(I)$, because
$r_m$ becomes significantly larger than $\sqrt{1/3}$, in particular
in the high spin region. The $I_m$-values for states in the wobbling
and the yrast band are not very different from each other and the
increasing negative value of $Q_0(I)$ is responsible for the difference
between $Q_2(I)$ and the full SQM. At high spins the contributions $Q_0(I)$
and $Q_2(I)$ are almost equal for large values of $I$. The difference of
the SQM between states in the wobbling and yrast band becomes
smaller, because $I_s$ and $I_l$ are almost constant for the
longitudinal alignment at high $I$-values and the denominator
$(I+1)(2I+3)$ causes the values in both bands to be small and close
to each other. At the same time $Q_0$ is responsible for the fact
that the absolute value of the SQM is not decreasing. For very high
$I$-values, one expects that due to the longitudinal alignment $I_m$
is dominant and approximately as big as $I$, such that $Q_2(I)$
vanishes and $Q_0(I)$ tends to $Q_0^\prime \cos\gamma$.

%%%%%%%%%%%%%%%%%%%%%%%%%%%%%%%%%%%%%%%%%%%%%%%%%%%%%%%%
%                    begin  summary
%%%%%%%%%%%%%%%%%%%%%%%%%%%%%%%%%%%%%%%%%%%%%%%%%%%%%%%%

In summary, the $g$-factor and static quadrupole moment (SQM) for
the wobbling mode of $^{135}$Pr, $^{105}$Pd, and $^{187}$Au have
been investigated in the framework of the PRM. The $g$-factor allows
one to deduce the expectation value of the rotor angular momentum
$\bar{R}$. Due to the negative gyromagnetic ratio of a
valence-neutron, the $g$-factor of $^{105}\mathrm{Pd}$ is increasing
with increasing spin $I$, in contrast to the decreasing $g$-factor
of the other two nuclides $^{135}$Pr and $^{187}$Au with a
valence-proton. The SQM depends on all three total angular
momentum component expectation values $I_s$, $I_m$, and $I_l$.
It is smaller in yrast band than in wobbling band for the
transverse wobblers $^{135}$Pr and $^{105}$Pd, while
larger for the longitudinal wobbler $^{187}$Au.

At present, there are no experimental data for the $g$-factor and SQM
of these three nuclei ($^{135}$Pr, $^{105}$Pd, and $^{187}$Au).
Therefore, we could only outline which information one can extract
from these observable and explain the tendencies of these quantities
in the PRM. Future experimental efforts along this direction are of
high importance. Moreover, theoretical investigations for $g$-factor
and SQM using more sophisticated methods, such as random phase approximation
based on cranking approach~\cite{Shimizu1995NPA, Matsuzaki2002PRC,
Shimizu2008PRC, Frauendorf2015PRC}, collective Hamiltonian based on
tilted axis cranking approach~\cite{Q.B.Chen2014PRC, Q.B.Chen2016PRC_v1},
or triaxial projected shell model~\cite{Sheikh2016PS,
Shimada2018PRC_v1, Y.K.Wang2020PLB}, will be interesting also.

\section*{Acknowledgements}

Financial support for this work was provided in parts by Deutsche
Forschungsgemeinschaft (DFG) and National Natural Science Foundation
of China (NSFC) through funds provided to the Sino-German CRC 110
``Symmetries and the Emergence of Structure in QCD'' (DFG
Project-ID~196253076, NSFC Grant No.~12070131001). The work of UGM
was also supported by the Chinese Academy of Sciences (CAS) through a
President's International Fellowship Initiative (PIFI)
(Grant No.~2018DM0034) and by the Volkswagen Stiftung
(Grant No.~93562).

%%%%%%%%%%%%%%%%%%%%%%%%%%%%%%%%%%%%%%%%%%%%%%%%%%%%%%%%
%                  begin refereee
%%%%%%%%%%%%%%%%%%%%%%%%%%%%%%%%%%%%%%%%%%%%%%%%%%%%%%%%

\end{document}